\documentclass[fleqn,usenatbib]{mnras}

\usepackage{newtxtext, newtxmath}

\usepackage[T1]{fontenc}

\DeclareRobustCommand{\VAN}[3]{#2}
\let\VANthebibliography\thebibliography
\def\thebibliography{\DeclareRobustCommand{\VAN}[3]{##3}\VANthebibliography}

\usepackage{graphicx}	% Including figure files
\usepackage{amsmath}	% Advanced maths commands
\usepackage[export]{adjustbox}
\usepackage{xcolor}

\usepackage{subfig}
\usepackage{subscript}
\usepackage{soul}

\newcommand{{\HII}}{H\,{\sc ii}}
\newcommand{{\HI}}{H\,{\sc i}}

\newcommand{\um}{\,$\mu$m}
\newcommand{\kms}{\,km\,s$^{-1}$}

\newcommand{\flux}{ erg s\textsuperscript{-1} cm\textsuperscript{-2} sr\textsuperscript{-1} cm}
\newcommand{\msun}{\text{M}\textsubscript{\(\odot\)}}
\newcommand{\lsun}{\text{L}\textsubscript{\(\odot\)}}

\title[Colliding Clusters in II Zw 40]{The Radio-Infrared Nebula in II Zw 40: Clusters Forming in Colliding Elongated Clouds }

\author[D. Beilis et al.]{Dan Beilis $^{1}$\thanks{danbeilis@mail.tau.ac.il}
Sara C. Beck,$^{1}$
John Lacy,$^{2}$
Jean L. Turner,$^{3}$
\newauthor Hauyu Baobab Liu,$^{4,5}$
Paul T.P. Ho,$^{6}$
S. Michelle Consiglio$^{3}$
\\
$^{1}$School of Physics and Astronomy, Tel Aviv University, Ramat Aviv, Israel 69978\\
$^{2}$Department of Astronomy, University of Texas, Austin Tx USA\\
$^{3}$ Department of Astronomy, UCLA, Los Angeles, Ca. USA  \\
$^{4}$Department of Physics, National Sun Yat-Sen University, No. 70, Lien-Hai Road, Kaohsiung City 80424, Taiwan, R.O.C.\\
$^{5}$Center of Astronomy and Gravitation, National Taiwan Normal University, Taipei 116, Taiwan\\
$^{6}$Academica Sinica Institute of Astronomy and Astrophysics, Taipei, Taiwan\\
}

\date{Accepted XXX. Received YYY; in original form ZZZ}

\pubyear{2024}

\begin{document}
	\label{firstpage}
	\pagerange{\pageref{firstpage}--\pageref{lastpage}}
	\maketitle
	\begin{abstract}
		 II Zw 40 is a starburst dwarf and merger product, and holds a radio-infrared supernebula excited by thousands of embedded OB stars.  We present here observations of three aspects of the supernebula:  maps of the K and KU radio continuum that trace dense ionized gas with spatial resolution  $\sim0.1^{\prime\prime}$, a spectral data cube of the [S IV]$10.5\mu$m emission line that measures the kinematics of the ionized gas with velocity resolution $4.5$\kms, and an ALMA spectral cube of the CO(3-2) line that probes the dense warm molecular gas with spatial and velocity resolution comparable to the ionized gas. The observations suggest that the supernebula is the overlap,collision or merger of two star clusters, each associated with an elongated molecular cloud.   We accordingly modelled the supernebula with simulations of colliding clusters. The model that best agrees with the data is a grazing collision that has distorted the gas and stars to create the distinctive structures observed.  These models may have wide applicability in the cluster-rich regions of young starbursts. 

	\end{abstract}
	\section{Introduction}  
	II Zw 40 is a peculiar blue compact dwarf galaxy 10.3 Mpc distant and the prototype 'Extragalactic \HII~ Region' (\citet{SearleSargent70},\citet{SearleSargent72}). It is believed to be the product of a merger between two dwarf galaxies and that the merger has triggered the starburst.  The optical emission is dominated by a giant \HII~ region  $\sim7\arcsec, 350$pc~ diameter holding many young star clusters.  The two brightest clusters are referred to as SSC-N and SSC-S or Sources A,B(\citet{KP14},\citet{van2008}). SSC-N  is the brighter by a factor of 20 and is the dominant energy source of the \HII~region.  The luminosity and mass of SSC-N deduced from optical and UV observations are an order of magnitude greater than those of 30 Doradus ($9.1\pm1\times10^5$\msun, $1.1\pm0.1\times10^9$\lsun, \citep{LBL2018}).   At longer wavelengths the central region of II Zw 40 is dominated by a radio-infrared supernebula excited by thousands of embedded O stars (\citet{BT02}, \citet{KP14}).  The spatial structure of the  emission is not simple; the 6-1.3cm observations of \citet{BT02} and \citet{KP14}  found multiple clumps which they interpreted as sub-clusters  but the clumps were at the edge of detectability, depended very much on the deconvolution parameters, and were not consistent between the various maps.  
	
As in many low-metallicity galaxies,  CO(1-0) in II Zw 40 is very weak and hard to observe \citep{Sage92}. The higher level CO lines are stronger; \citet{Con16}  show high-resolution CO(3-2) maps made with ALMA,  and \citet{KP16} report CO(3-2) and (2-1) maps.  The CO is concentrated into small clouds, one of which coincides with the radio-infrared source.   \citet{KP16} find clumps at multiple velocities over a total range of $\approx30$\kms, with most of the molecular emission at $750\pm10$\kms, close to the galactic velocity of 773\kms (derived from stars, \citet{Mould2005}; the HI velocity is 800\kms, \citet{Spring2005}).  The small CO clouds are not seen to be directly related to the atomic and molecular structures seen at larger scales. 
	
	Optical and near-IR observations of shocked and ionized gas in the starburst region find  gas motions strongly affected by the energetic activity of the young stars (\citet{Bordalo2009});\citet{van2008} do not find any connection to the dynamics and structure of the galaxy on larger scales.   There are convincing arguments that the starburst was triggered by cloud-cloud collisions as the two dwarf galaxies merged; the tidal tails observed in HI \citep{vanZee95} are witness to this history.  On scales of the starburst, winds and outflows of the young stars have muted traces of these interactions.  
	
	Observations of the central $\sim300$~pc of II Zw 40, as summarized above, present a picture of a region where star clusters have very recently been born.  In this paper we attempt to  deduce the history of the current active starburst (that centered on the radio-infrared supernebula).  We address this question from two angles. We first present new high spectral and spatial resolution observations of the dense and compact components of the ionized gas.  This gas is directly associated with the bright, embedded and young star clusters rather than with the giant \HII~region.  The new data probes the density distribution of gas and stars inside the clusters as well as the motion of the gas ionized by the young stars.  We combine this new data with CO(3-2) ALMA observations of the molecular gas in which the youngest clusters are embedded. This data reveals the immediate kinematic environment of the starburst and the molecular gas which fuels it.   The observations are detailed in Section 2, the spatial structure of the ionized gas and molecular gas in Section 3, and the kinematics of  ionized and molecular gas in section 4.  The observations motivate the second aspect of the paper, wherein we model the II Zw 40 starburst as the collision of two star clusters which have formed in two elongated molecular clouds.   We have run simulations of colliding and interacting molecular clumps and of star clusters merging under a wide range of conditions, which are discussed in Section 5. In Section 6 we compare our simulations to the observations, arrive at a favoured model, and predict the future development of the II Zw 40 clusters. 
	\section{Observations and Data Reduction}
	\subsection{Radio Continuum Maps}
	II Zw 40 was observed in JVLA  project 16B-067.   The K-band (22 GHz) observations were carried out on 15 October 2016 and Ku (15 GHz) on 31 October 2016. 
Total execution times were 00:43:06 and 00:54:16, and the bandpasses 6 GHz, 4GHz respectively.   The data was handled with  the standard JVLA pipeline process \footnote{https://science.nrao.edu/facilities/vla/data-processing/pipeline}, version 4.7.1,  and since in this time period the atmospheric delay model was not applied properly \footnote{https://science.nrao.edu/facilities/vla/data-processing/vla-atmospheric-delay-problem} all data from affected observations were re-processed with a corrected model.  
The data were cleaned and imaged with CASA and AIPS. The galaxy was mapped with natural weighting for greatest sensitivity and Briggs weighting (robust = 0) for higher resolution.  The observational parameters are given in Table 1.    
	  
	\subsection{Mid-Infrared Spectroscopy with TEXES}
	II Zw 40 has low metallicity and high excitation \citep{LBL2018} and its mid-infrared spectrum, as seen by the Spitzer IRS \citep{wu06}, is dominated by the high excitation lines of [NeIII] 15.5\um~ and [S IV] 10.5\um.  [NeIII] cannot be observed from the ground so [S IV] is used to probe the ionized gas kinematics \citep{BT13}.  [S IV] has two great advantages over \HI~ recombination lines; it is less affected by extinction than many, including Br$\gamma$, and it is much less affected by thermal broadening than any hydrogen line can be.  Thermal broadening in an \HII~ region at $T_\text{e}\sim10^4$K will give any hydrogen line FWHM about 20\kms~, while a line of a metal of mass $m_\text{i}$ will have width $\sqrt{\frac{m_\text{H}}{m_\text{i}}} \times$ that of  hydrogen.  
	
	We accordingly observed [S IV] in 
	II Zw 40 with the TEXES spectrometer on Gemini North on 15 March 2017. TEXES, the Texas Echelon Cross Echelle Spectrograph \citep{LRG2002}, operates between 4.5 and 25 \um~and has spectral resolution $R\approx 4000-100,000$.  These observations were performed in high resolution mode, using a 32 line/mm echelle and a 0.5\arcsec~ wide, 4\arcsec~ long slit.  The $S^{+++}$ thermal broadening of $\approx 3.5$\kms~, added in quadrature to the instrumental resolution,  gives the true resolution of $\approx4.5$\kms~.   
	The asteroid Vesta was the calibrator. The slit was aligned north-south and stepped east-west across the targets and the scan combined into a data cube of 23  0.25\arcsec~ pixels in R.A.,  32 0.14\arcsec~ in Dec, and  256  0.95\kms~ wide velocity channels.     
	\subsection{Archival ALMA Observations of Molecular Gas}
	II Zw 40 was observed in ALMA Band 7, 345 GHz, 
	as a Cycle 2 (Early Science)
	program (ID = 2013.1.00122.S). Band 7 observations took place on 2014 13 August and 13 December; weather
	conditions were good, with $ T_\text{sys}=140$-250~K and 
	PWV$\sim$0.78mm.
	Observations of a single field with an 18\arcsec\ field-of-view centered on
	05:55:42.620 +03.23.32.0 (no offset)
	were  concatenated with total time 2844 seconds on source.  Bandpass and phase were calibrated
	with J0607-0834 and J0532+0732 respectively; J0510+180 was the flux calibrator.
	Calibration of the data was done in the pipeline\footnote{https://almascience.eso.org/processing/science-pipeline} and with CASA 4.2.2. 
	Continuum emission was subtracted in the (u,v) plane before making line maps. 
	
	\section{Spatial Distribution of Molecular and Ionized Gas in the Starburst }
	\subsection{Radio Continuum: Multiple Clusters?}
	
	\begin{figure*}
		\includegraphics*[width=0.75\textwidth]{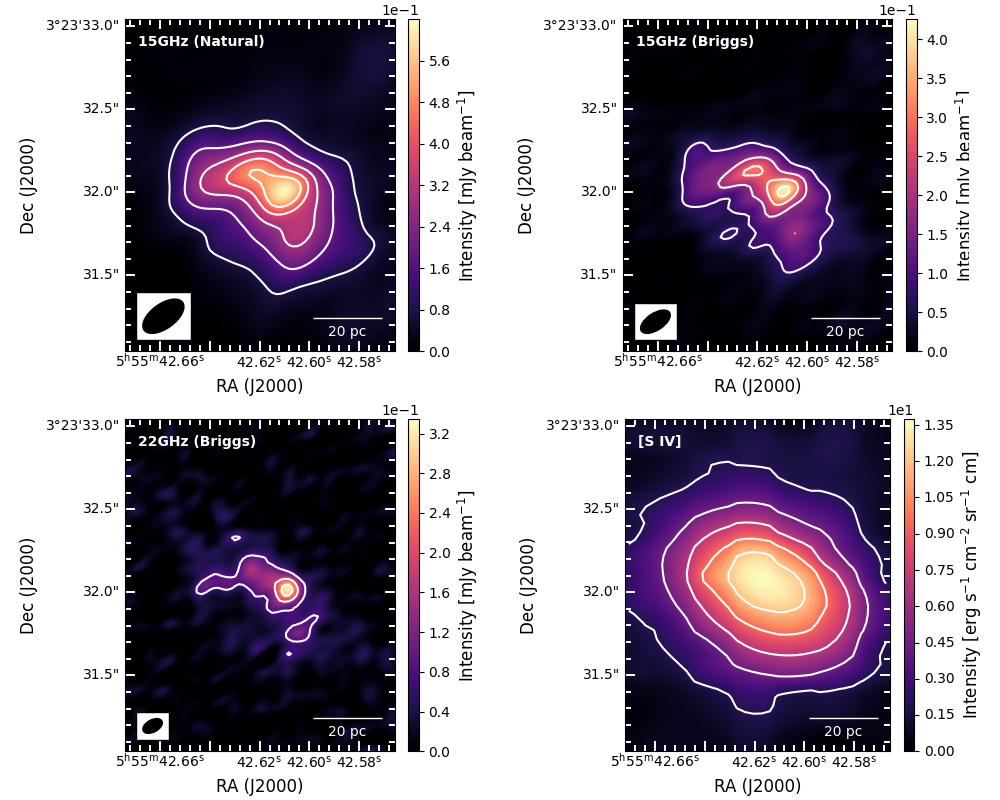}
		\caption{II Zw 40 imaged at (clockwise from upper left):  a) 15 GHz (natural weight), b) 15 GHz (Briggs weight), c) 22 GHz (Briggs weight), d) [S IV] 10.5\um m, 0.3'' beam.  The intensity wedges have units mJy beam$^{-1}$ for the radio and \flux~for the [S IV] and the contours levels are [0.8,1.85,2.9,3.95,5] for the radio maps and  [2,4.5,7,9.5,12] for the [S IV].    } 
		\label{fig:siv_radio}
	\end{figure*} 
	
	Figures \ref{fig:siv_radio}a+b and \ref{fig:siv_radio}c  show II Zw 40 at 15 and 22 GHz.  The source has an extended emission envelope with some clumpy structure,  dominated by a  bright source at RA 05:55:42.60  DEC 03:23:32.0. This position agrees with \citet{KP14}'s Source D and with their position for SSC-N.  It is the only distinct emission peak.     The highest resolution radio map, Figure \ref{fig:siv_radio}c,  has a  0.12\arcsec~ beam and shows the bright clump to be resolved with size 0.13\arcsec~  or $\approx6.5$ pc (in both axes) and flux 0.59 mJy.    The JVLA maps, which are deeper than those of \citet{BT02} and have noise lower by a factor $\approx5$, show a rather different picture of the embedded HII regions.  Where  \citet{BT02} saw $\approx2\sigma$ brightness fluctuations as possible secondary clusters and could not detect most of the extended emission, the current data show that ionized gas has an unusual and asymmetric extended structure and that the spatial distribution is dominated by one large cluster. 
	
	\subsubsection{The [S IV] Agrees with the Radio Continuum}
	Figure \ref{fig:siv_radio}d shows the [S IV] emission summed over the entire grating order, covering velocities from 600 to 880\kms~.  The TEXES beam size at Gemini was $\sim0.3\arcsec$, close to the diffraction limit; the pixel size, set by the scan method, was $0.25\times0.144\arcsec$.  Comparing the [S IV] map to the 15 GHz radio map of Figure \ref{fig:siv_radio}a, which has close to the same resolution, shows very similar spatial distribution.  This confirms that the [S IV] emission is closely associated with the embedded star cluster which excites the radio continuum peak, and is not affected by emission of the larger optically bright \HII~ region.

	\subsection{Spatial Structure of Dense Molecular Gas}
	
\begin{figure*}
		\begin{center}
			\subfloat[][]{\includegraphics[width=0.375\textwidth]{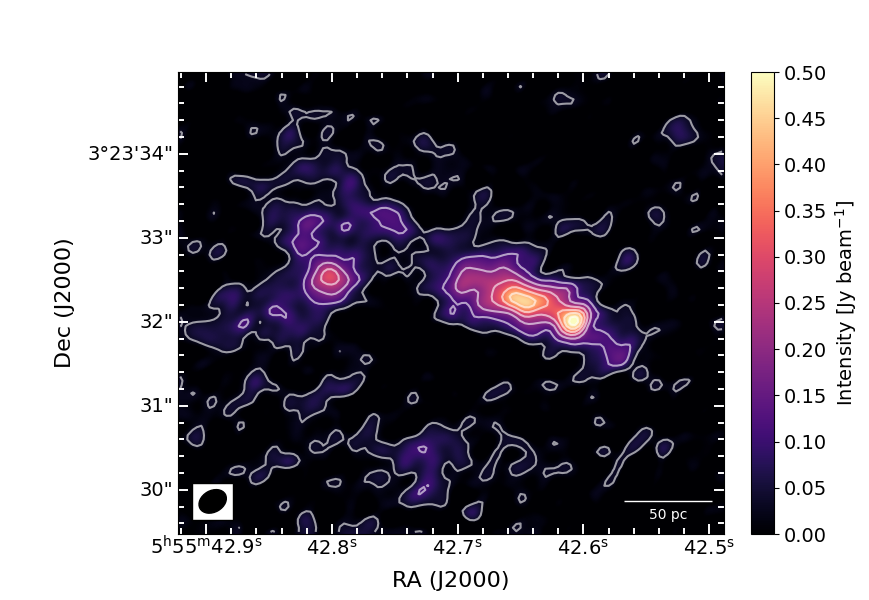}}
			\subfloat[][]{\includegraphics[width=0.375\textwidth]{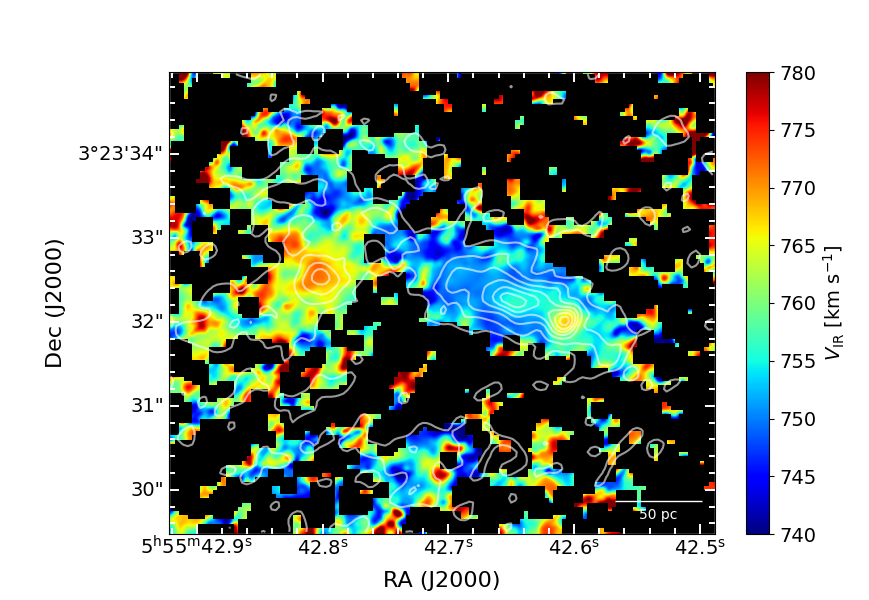}}
			\hfill
			\subfloat[][]{\includegraphics[width=0.375\textwidth]{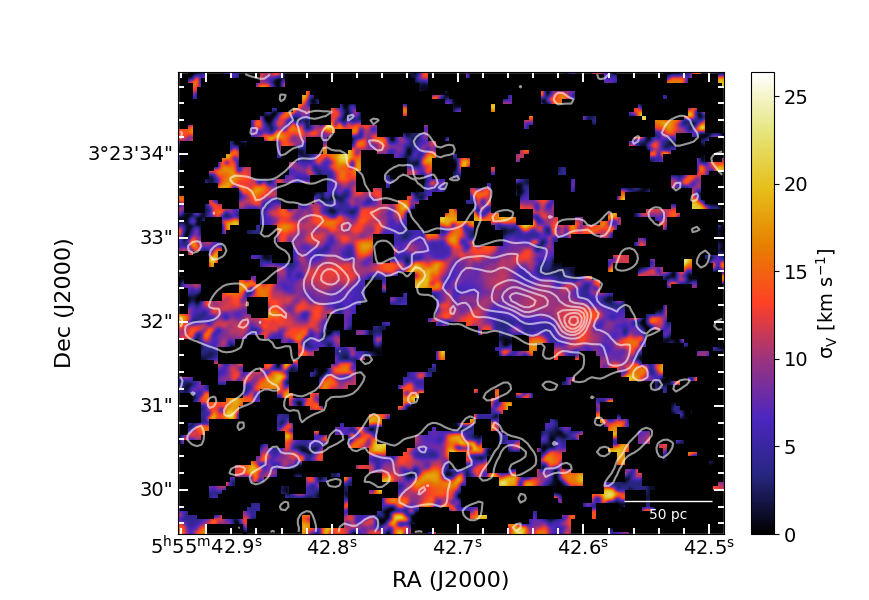}}
			\subfloat[][]{\includegraphics[width=0.375\textwidth]{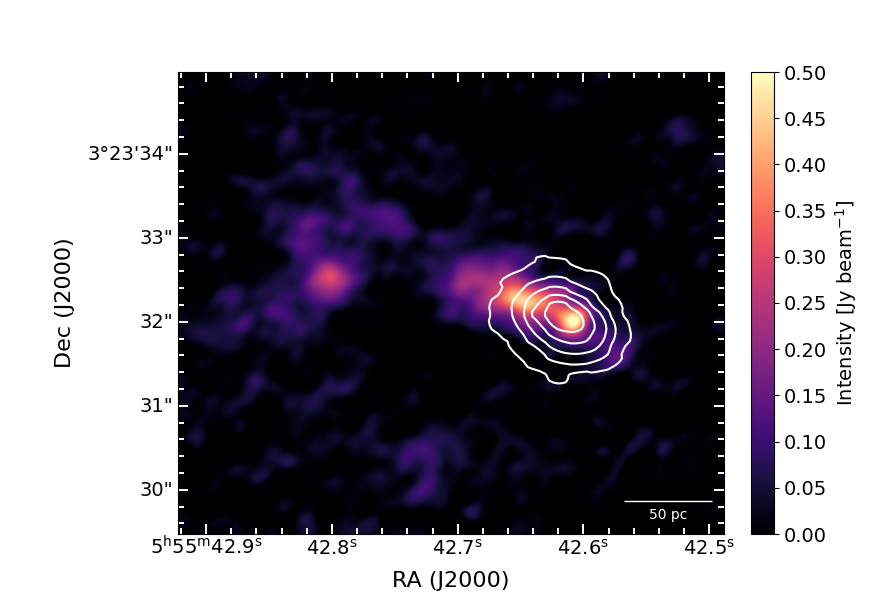}}
			\caption{II Zw 40 SSC-N CO(3-2) (a) 0\textsuperscript{th} moment map with contours at [0.03, 0.11, 0.19, 0.26, 0.34, 0.42, 0.5] Jy beam$^{-1}$, (b) 1\textsuperscript{st} moment map, (c) 2\textsuperscript{nd} moment map and (d) 0\textsuperscript{th} moment map with [S IV] contours at [2,4.5,7,9.5,12]\flux. The maps in panels a,b, and c are overlayed with the CO 0\textsuperscript{th} moment map contours.}
			\label{fig:co_moments}
		\end{center}
	\end{figure*}

	Figure \ref{fig:co_moments} displays the first 3 moments of the CO(3-2) data cube (integrated line intensity, intensity-weighted velocity, and velocity dispersion). The 4th panel of the figure shows the integrated [S IV] line intensity overlaid on the molecular gas distribution. The molecular gas has larger extent than does the ionized and is distributed in several distinct density peaks.  The dominant molecular peak  marked as 'W4' in Figure \ref{fig:co_profiles}, agrees with 
	\citet{2017ApJ...850...54C}'s CO and 3 mm continuum peak 'W',  and is co-incident with the radio and infrared cluster.   NE of this peak emission of the ionized gas drops off sharply at the border of \citet{2017ApJ...850...54C}'s  component 'C' ('W3' in  Figure \ref{fig:co_profiles}).   The spatial distribution of the high-resolution Briggs radio map suggests that the radio emission is blocked from extending into component 'C'; if real and not an artifact of the display, this may show a high density ionization front.  We do not detect ionized gas in \citet{2017ApJ...850...54C}'s source 'E' ('E' in \ref{fig:co_profiles}).
	
	\section{Gas Kinematics in the Starburst}
	\subsection{Ionized Gas:Double Peak or Extended Wing?}
	\begin{figure}
		\includegraphics[width=1\columnwidth]{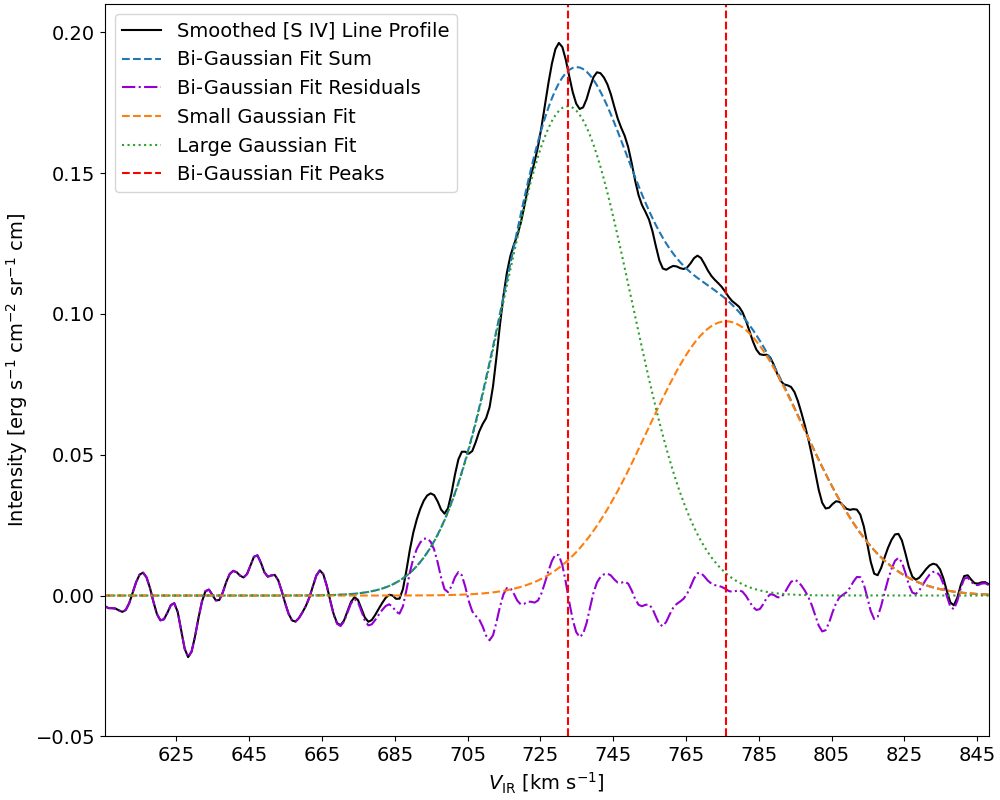}
		\caption{[S IV] emission summed over entire source, smoothed by 3 pixels (1  resolution element).  X axis units are \kms~ and Y axis units are  \flux}
		\label{fig:siv_lineprofile}
	\end{figure}

	\citet{BT13} observed II Zw 40 with TEXES at the NASA IRTF and obtained spectra with  1.2\arcsec~ spatial resolution.  They found that the [S IV]  line emission extended to high velocities redwards of the peak and that the line profile could be fit by two gaussians separated in velocity by  $\approx44$\kms~.  \citet{BT13} accordingly suggested  that two super star clusters offset in velocity were included in the same beam, a model which motivated the higher resolution observations of Gemini. 
	
	Figure \ref{fig:siv_lineprofile} shows the Gemini [S IV] line profile summed over the entire source.  The new data confirm that 1) the line peaks at $745\pm5$\kms~, 2) emission covers an extraordinarily large velocity range of $\sim130$\kms~ FWZI, and 3)  the line profile is extremely asymmetric, extending to $\sim100$\kms~red and only  $\sim50$\kms~ blue from the peak with excess emission on the red side of the line peak. The line profile could be formally fit by 2 gaussians at $\sim744, 775$\kms~ consistent with the IRTF results \footnote{The [S IV]  velocities given in \citet{BT13} differ from those here because they were on a geocentric, not heliocentric, system.}
,  although it does not rule out large scale gas flows in addition to the two peaks.   
	
	The simplest interpretation of the line profile is that the source holds two clusters, the larger $\approx2$ times the brightness of the smaller.  If this is so, can we see any sign of the second cluster  in the spatial distribution of the ionized gas?    If the red excess is
	associated with one of the sub-clumps seen in the high resolution JVLA and ALMA maps it would 
	support the model of two off-set clusters.   
	
\begin{figure*}
		\begin{center}
			\includegraphics*[width=0.75\textwidth]{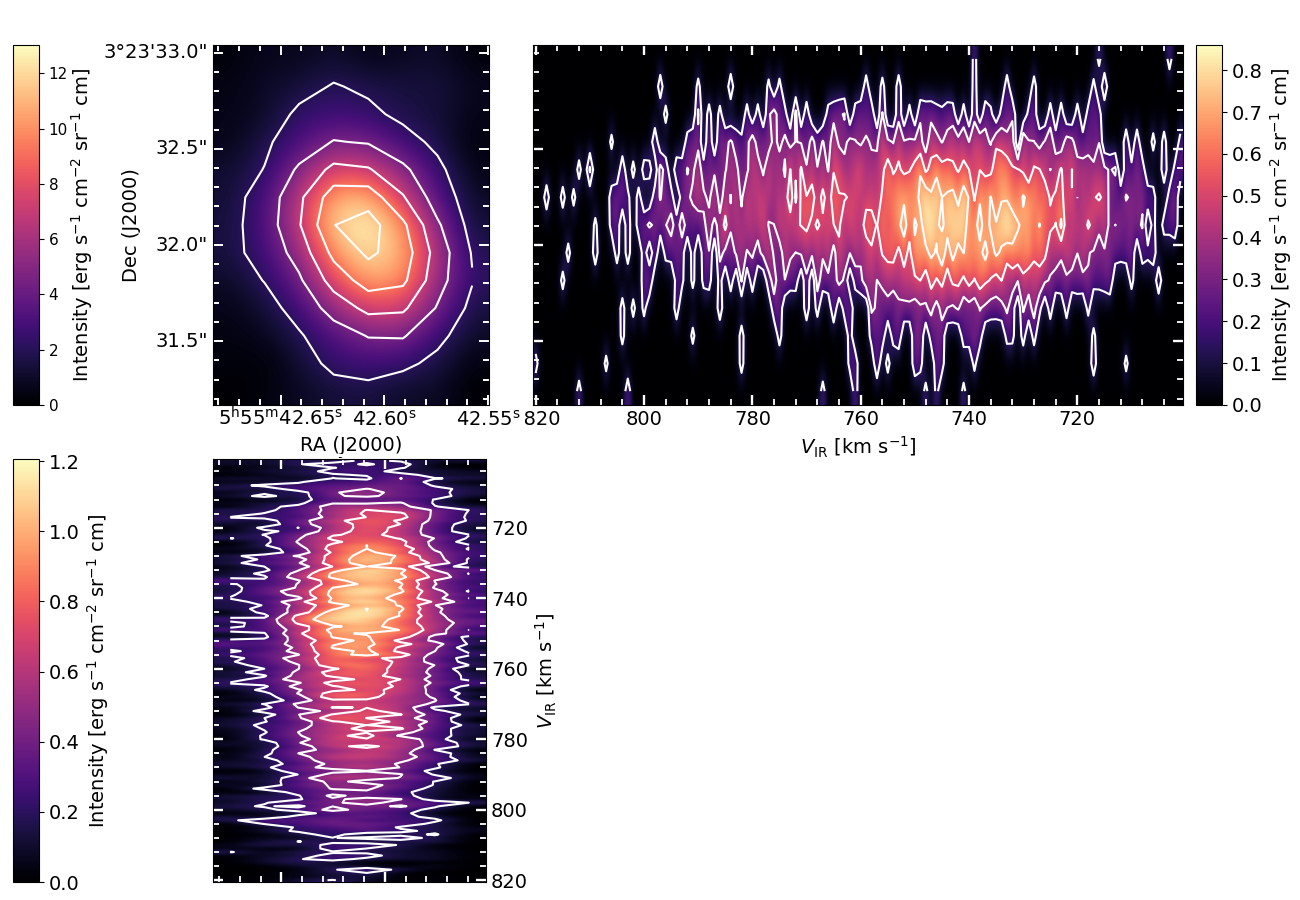}
			\caption{The [S IV] data cube collapsed along the axis of velocity (top left), R.A. (top right) and Dec. (bottom).   Emission redder than 
			$\approx745$\kms~ is slightly asymmetric in Declination, extending further south. The contours for the top, right and bottom subfigures are respectively [2,4.5,7,9.5,12], [0.15,0.34,0.53,0.71,0.9] and [0.2,0.45,0.7,0.95,1.2]\flux }
			\label{fig:siv_collapse}
		\end{center}
	\end{figure*}

	 Figure \ref{fig:siv_collapse} displays the [S IV] data cube collapsed along the velocity (moment 0), R.A. and Declination axes. Both  components are 
	extended.   Examining the line profiles in each pixel (as in the Appendix figure) confirms that both features appear in every pixel and that the blue  is the stronger except for a small area of the extreme south-east where the red dominates.  But in none of the radio or infrared maps of the ionized gas is there a spatial density peak or clump associated with the red emission. 
		
	We conclude that if there are two distinct clusters, as the [S IV] profile strongly suggests,  they so overlap in the line of sight that they cannot be separated.

	\subsection{Molecular Gas: Two Elongated Clouds at Different Velocities}
	\begin{figure*}
		\begin{center}
			\includegraphics*[width=0.75\textwidth]{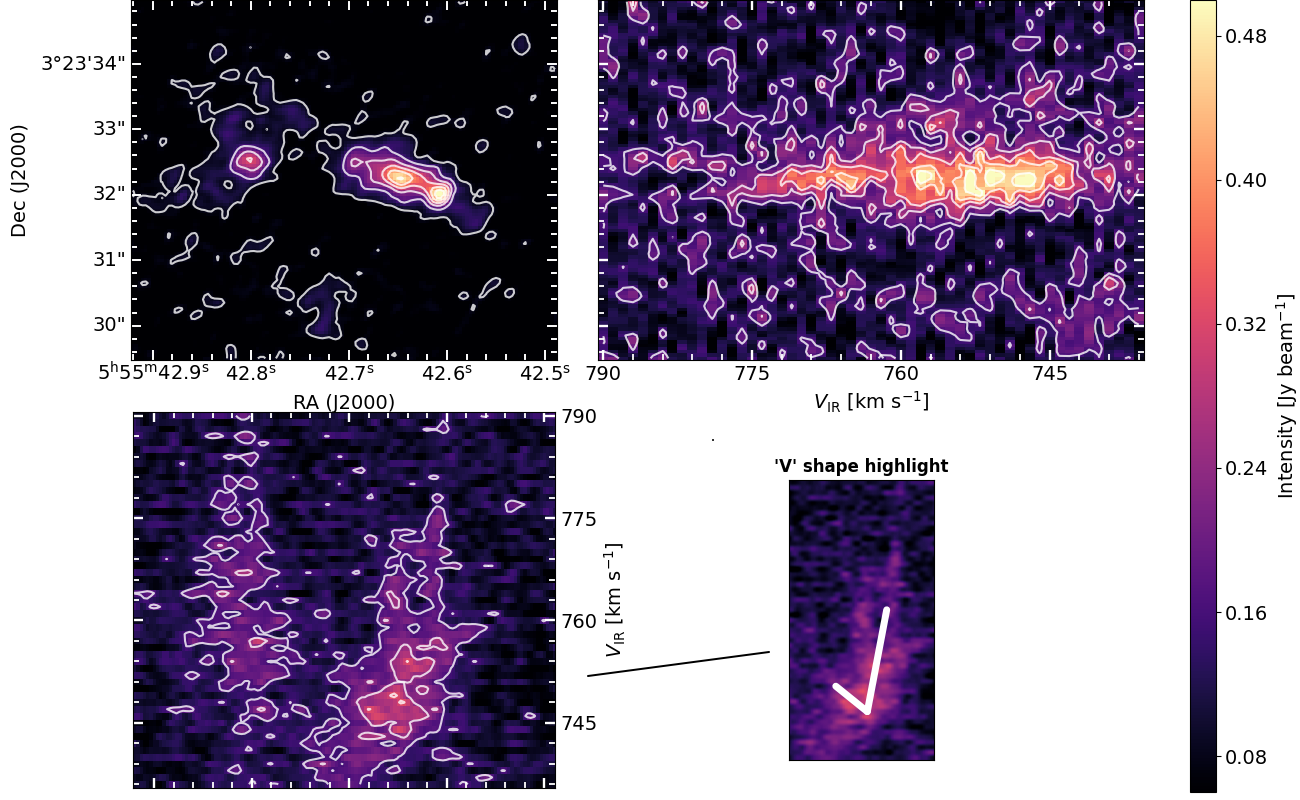}
			\caption{The CO(3-2) data cube collapsed along the axis of velocity (top left), R.A. (top right) and Dec. (bottom left). The contours are at [0.08, 0.16, 0.24, 0.32, 0.4, 0.48] Jy beam$^{-1}$.   (bottom right) The western cloud, indicated by the pointer line in the previous figure,  expanded and marked in white to highlight the potential observational signature of a cloud-cloud collision \citet{Torii2017}. }
			\label{fig:co_collapse}
		\end{center}
	\end{figure*}
	
\begin{figure}
		\includegraphics*[width=1\columnwidth]{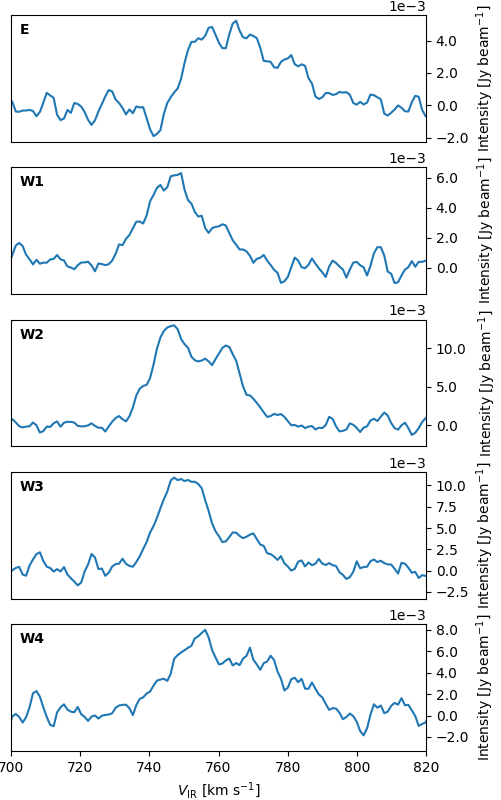}\\
			\includegraphics*[width=1\columnwidth]{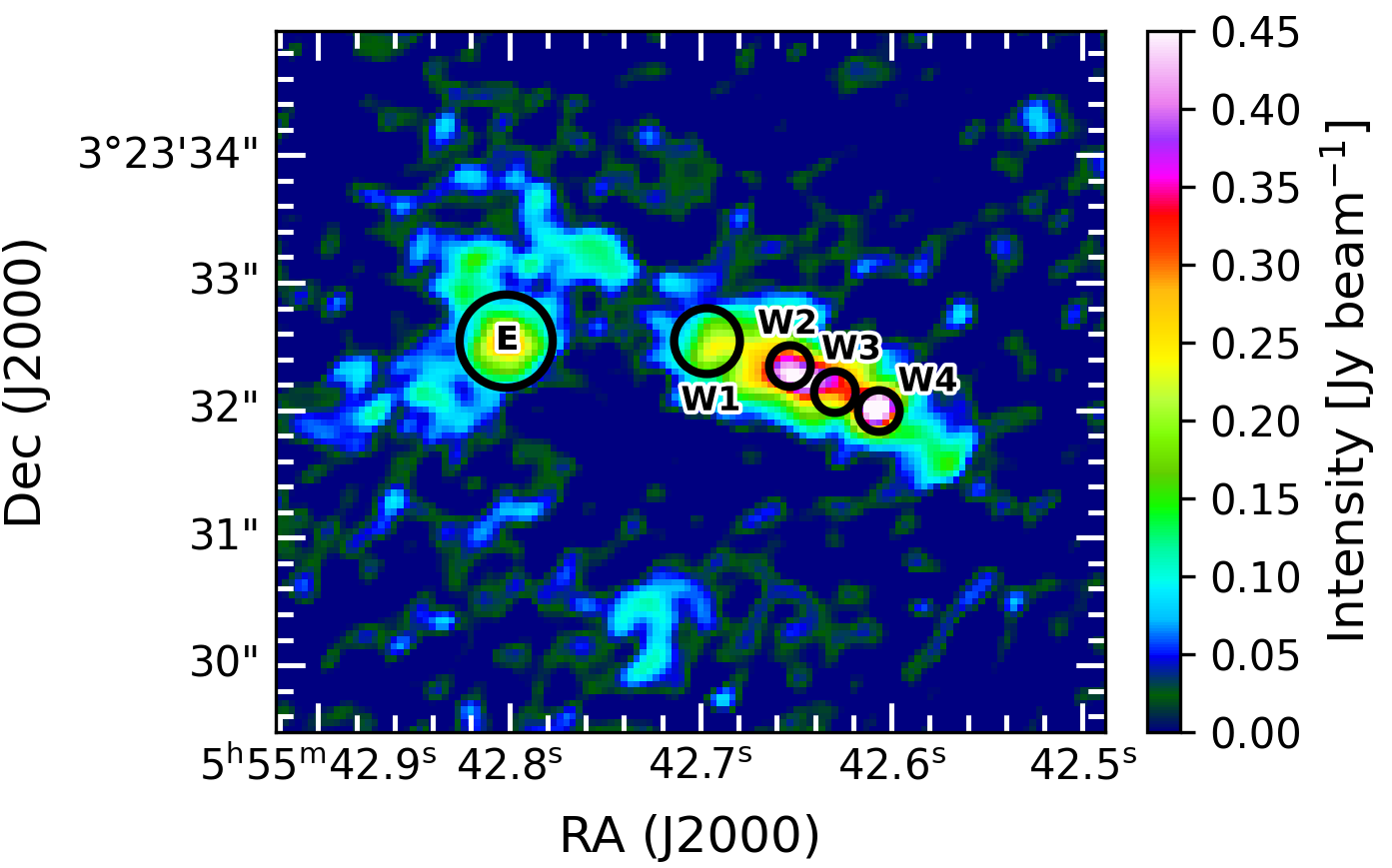}
		\caption{Top, line profiles of CO(3-2) in 5 positions across the source. Bottom, the MOM0 map of CO(3-2) with circles identifying the location and showing the area over which the data was summed for each line profile.		\label{fig:co_profiles}}
	\end{figure}

	Figure \ref{fig:co_collapse} shows the CO data cube collapsed along the velocity, declination and right ascension axes, and together with Figure \ref{fig:co_moments} demonstrates  that both the intensity-weighted velocity and the dispersion of the CO line are larger going from east to west along the streamer and reach maxima on the radio source.  The line profile (shown in Figure \ref{fig:co_profiles}) also varies markedly over this region.    In the eastern clump (Feature E of \citet{2017ApJ...850...54C}) the profile is flat-topped, almost rectangular, and covers  $\sim750-780$\kms~.  In the main molecular streamer that includes the radio and [S IV] sources, two velocity features appear along the length of the emission.  The blue peak is consistently at $\sim745-750$\kms~ and the red varies between  $\sim760-772$\kms~; the profile on the radio-[S IV] source resembles that of the [S IV] line but does not extend quite as far into the red.  (These velocity features can be seen also in the channel maps in the Appendix).
	
	The Position-Velocity diagram (PVD) along the Dec axis resembles the V-shape observational signature of a cloud-cloud collision \citet{Torii2017} (highlighted in a subfigure in Figure \ref{fig:co_collapse}). This shape would usually be skewed when observed at an inclination angle. That the Dec PVD appears closer to a check mark than a 'V' may indicate that a more complex collision is occurring.

\begin{figure*}
		\begin{center}
			\includegraphics[width=0.7\textwidth]{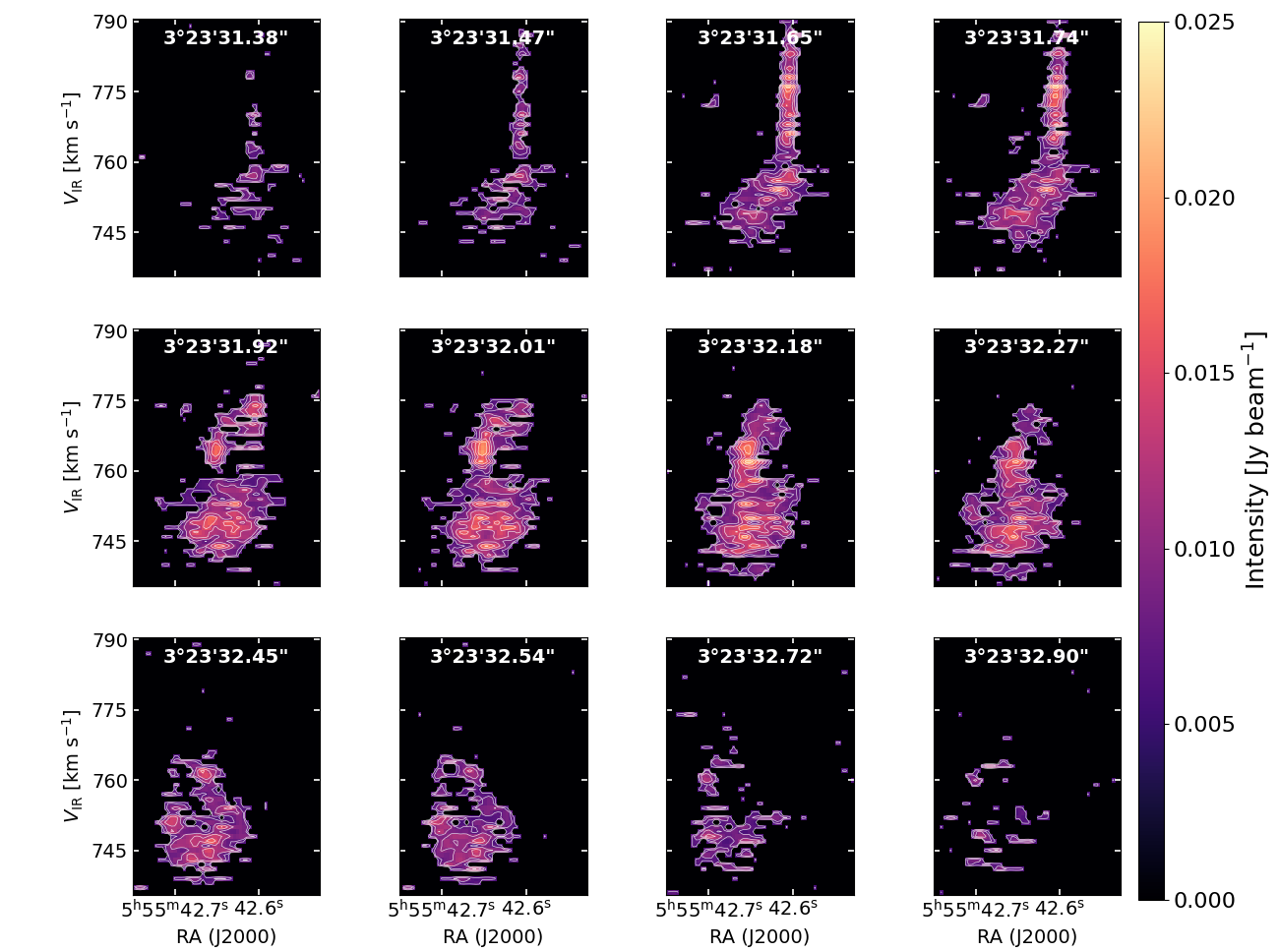}  \\
			\includegraphics[width=0.7\textwidth]{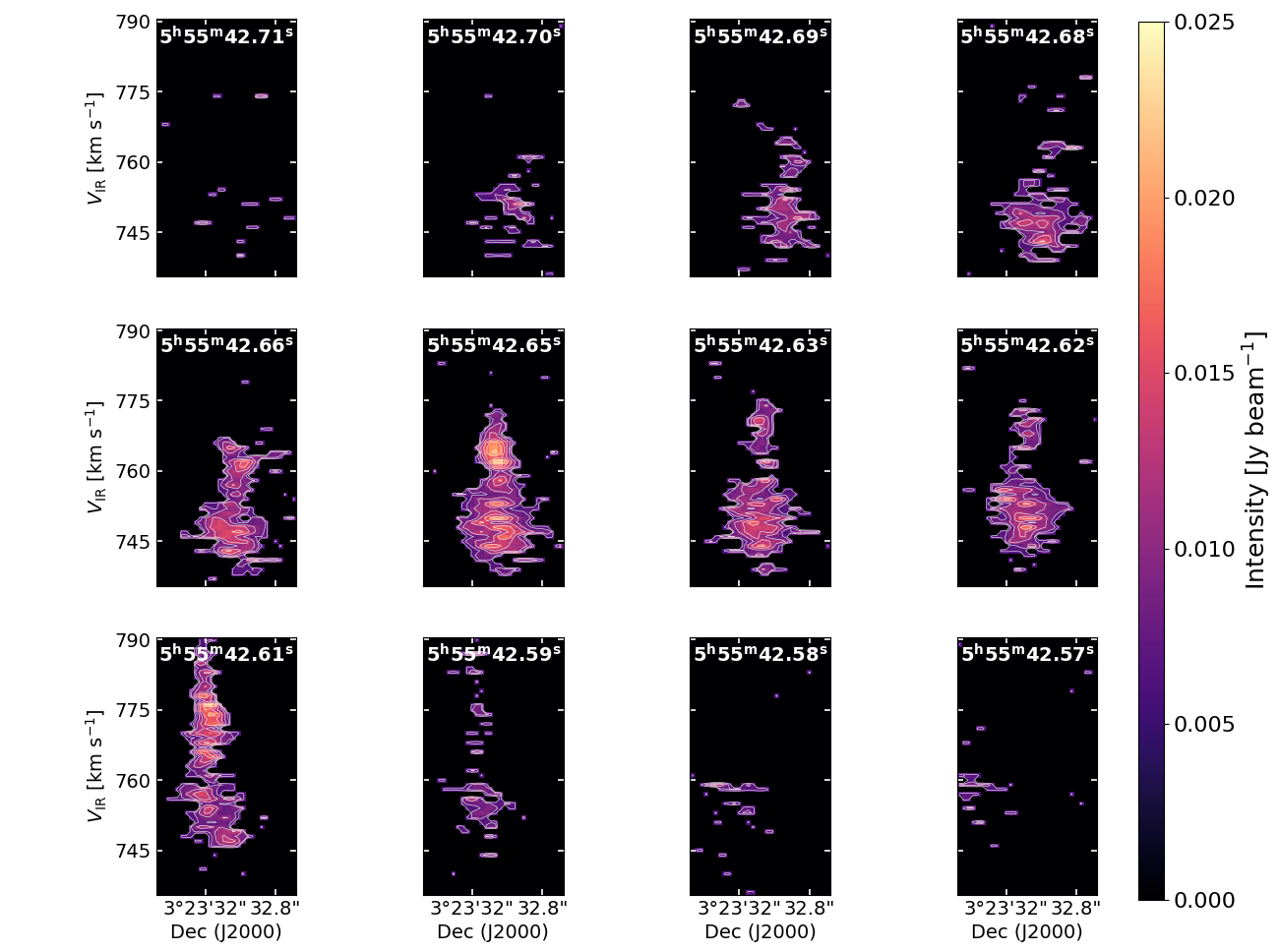}\\
			\caption{Position-Velocity Diagrams of the CO(3-2) observations along the Dec axis (top) and the RA axis (bottom). The contours are at [0.005, 0.0075, 0.01, 0.0125, 0.015, 0.0175, 0.02] Jy beam$^{-1}$.}
			\label{fig:co_pvd}
		\end{center}
	\end{figure*}

	More insight into this complicated velocity field may be found in the Position-Velocity Diagrams (PVDs) of Figure \ref{fig:co_pvd} (As the molecular source E is not associated with embedded star formation it will not be included in the rest of this discussion).  The PVDs are consistent with two elongated clouds overlapping in the line of sight, consistent with the double line peaks of Figure \ref{fig:co_collapse}. This motivates the working picture of II Zw 40 which we will explore with the simulations of the next section.   \footnote{Note that the CO and [S IV]  velocities do not agree with the $H\alpha$ velocities observed by \citet{BV09}, which are close to 780\kms~ over the entire emission region; the optical velocity field is highly complex and dominated by large expanding bubbles.}
	
	\section{Simulations: Colliding/Merging Clusters and Clumps}
	The infrared, radio, and sub-millimeter  observations reported combine to show that  the II Zw 40 core holds two elongated molecular clouds that overlap in the line of sight, and to suggest strongly that  two star clusters at different velocities, also overlapping in the line of sight,are present. The active 
	star formation region with the embedded star cluster(s) is at one end of the spatially overlapping elongated molecular clouds, where the line profiles of the clouds  merge to form a single broad feature.   These observations strongly suggest that the clouds are colliding or merging on the west end and that the star formation 
	is concentrated in the collision region.  This has motivated us to model the II Zw 40 starburst with simulations of colliding clusters.
	
	Theoretical calculations and models (extensive work including \citet{2018takahira}, \citet{1992Habe} and references therein)  have established  that the increased gas pressure in colliding clouds can increase or trigger star formation.   Such collisions are probably the main mechanism of the starbursts in interacting galaxies.  It has however been very difficult to observe collision-induced star cluster
	formation directly.   As soon as the young stars turn on, their  winds and radiation start to disperse the surrounding clouds, so there is only a short  ($\sim2$Myr, \citet{Fukui2016}) time in which the collision and star formation 
	processes can be observed.  We are fortunate that II Zw 40 appears to be now at exactly the stage where massive young stars have formed but enough of the natal clouds remain
	that we can model their interaction.  We accordingly have developed code to simulate cluster formation in clump collision and use it to reproduce as well as possible the observations of II Zw 40. 
	In the next sections we briefly describe the numerical methods, the results, and the model we believe to best explain II Zw 40. 
	
	\subsection{Numerical Methods: the AMUSE Environment}
	A full picture of the embedded clusters in II Zw 40 must consider hydrodynamics of the gas,  gravitational interactions of the stars and of the stars with the gas,  the evolution of the embedded stars and the effects of their winds (and ideally the magnetic field, for which there is at present no data).  The Astrophysical Multipurpose Software Environment (AMUSE) assembles community codes and makes it possible to run computations of all these aspects in parallel and to communicate the results between codes.  AMUSE couples all the codes, N-body and hydrodynamical, via a Bridge scheme \citep{Fujii2007}.  For these simulations we used the community codes ph4 \citep{Dolcetta2011} for the N-body calculations and GADGET-2 \citep{Springel2005} for Tree-SPH calculations.  We follow the evolution of the cluster stars with the parameterized stellar evolution module SeBA \citep{Zwart2018}, which gives stellar radius, luminosity and temperature as functions of time.  SeBA returns a mass loss rate from stellar winds which, input to the AMUSE stellar winds module \citep{vanderhelm2019}, generates more particles for the SPH calculations.  Stellar winds are expected to develop very differently  in dense embedded clusters than in the field \citep{wunsch2011}; we accordingly
	use  the stellar wind module's 'heat' mode, which follows the energy rather than the velocity of the wind particles, as most appropriate for this embedded source. 	
	\subsection{Simulations} 
	\subsubsection{Cluster Evolution}
Our observations show that the great cluster SSC-N coincides with the spatial overlap of two elongated molecular clouds, and establish the line-of-sight velocity offset of the clouds.  This motivates us to model SSC-N as the result of the cloud's interaction.  In this paper we start with the assumption that a star cluster has already formed in each clump, and follow the appearance of the resulting system through the collision. (We intend in future work 
to examine separately the alternate case of star formation starting after the collision).  The collision thus starts with physically realistic clusters,  rather than with an assumed density and velocity distribution.   We estimate from the line profile that the larger cluster is roughly twice the mass of the smaller, and from the total ionizing flux $N_\text{Lyc}$ of SSC-N that the total populations of massive O stars are approximately 4300, 2150 O stars respectively. The model clusters therefore have total masses of $4.5\times10^5$\msun~ and $2.2\times10^5$\msun where we assumed the gas comprises 70\% of the total mass  as the star formation efficiency in SSCs is often quite high \citep{lada2003embedded}.   
	 
	The initial conditions were chosen to be typical of stellar clusters: for the gas we assumed an idealized Plummer sphere \citet{plummer1911problem}, a Maxwell-Boltzman velocity distribution matching a temperature of $T=10^4$K and a particle mass of $1.5$\msun. For the stars, we assumed a King model \citet{king1966structure} and a Salpeter initial mass function \citet{salpeter1955luminosity} in a mass range of $[15,120]$\msun.  We chose these 'top heavy' mass limits to keep the total stellar mass and the computational demands realistic;  the winds of low mass stars are so much weaker than those of massive stars that they can be omitted with little effect on the large-scale results.      The clusters were allowed to evolve for a total of 2Myr with a 10kyr timestep. We chose the total evolution time based on the maximum dynamical relaxation time of the gas for both clusters based on their mass, number of particles and virial radius \citet{spitzer1987dynamical}. In each timestep, the stellar evolution module was allowed to evolve for half a timestep each time as it required a higher temporal resolution, then the bridge code (combining the gravitational and hydrodynamical code modules)  and stellar wind module was allowed to evolve for a full timestep and finally the stellar evolution module was allowed to evolve for an additional half timestep. After each code module ran, the appropriately affected parameters were copied from one code module to the other using the AMUSE inter-module communication scheme.
	
	After the evolution of each cluster, we examined the rate of mass loss (SPH particles blown away due to the stellar winds), radius, mass and energy to make sure that they were behaving in a typical manner of stellar clusters. This also helped us determine the best AMUSE modules for our purposes and what settings to run them at; The stellar wind module was set to the 'heating' kind based on the results of \citet{vanderhelm2019} and confirmed by examining the radius and number of SPH particles blown away by each stellar wind mode. Additionally, the mass of the SPH particles was chosen by running the cluster evolutions using different masses and determining which one provided the sufficient accuracy for our case. 
	\subsubsection{Cluster Collision}
	The  simulations of cluster collisions were run exactly the same way as the cluster evolutions, including both stellar evolution and stellar winds. The clusters were then trimmed to a little over the virial radius of each (which was also higher than the half-mass radius). This resulted in clusters that still had a large percent of the original gas while also preserving the core and outer layer structure. The primary reason for doing so was to reduce the run-time of the collision simulations while still maintaining sufficient approximations of realistic cluster structures and gas and stellar mass order of magnitudes estimated for the colliding clusters in II Zw 40.   In addition, the main focus in these simulations was to find a match to the two velocity peaks,% as seen in theW2-W4 of Figure \ref{fig:co_profiles}: 
	to which the cores of the colliding clusters are the most important contributions.  
	
	For the actual collision, the large cluster was placed at the center of the coordinates system and the small cluster was set some distance away with initial velocity of approximately $120$\kms. This velocity was determined by running multiple low-scale simulations, approximating the resulting line profile, and comparing it to the [S IV] line profile. The small cluster
	was then set to move at an angle so as to result in an approximately horizontal distribution of the two clusters relative to each other after the collision (see Figure \ref{fig:collision_density_comparison}). The impact parameter $b$, defined as the distance between the cluster centers, is crucial in determining the outcome of the collision.   We ran simulations for a range of $b$ starting from 0, the head-on collision, and increasing by 0.5 pc increments to 6pc, where the cluster interaction is minimal.  
	
	The time step is the same as in the evolution simulations, $10^4$yr, and the total runtime for each collision was 0.5Myr. After all the collisions were finished, the AMUSE results were used as input for the radiative transfer software RADMC-3D \citet{dullemond2012} to produce 3D line profiles. For the stars this included their mass, radius, temperature and position, and for the gas their mass, velocity and position. The properties of the SPH gas particles were interpolated to a grid of size 256x256x256 that covered the entire range of gas distribution which was 60x60x60 pc. Each collision was observed at an inclination angle of 45, 60 and 75°, as these angles gave a better line profiles match to the original [S IV] observations. The 3D line profile simulations were made in both CO(3-2) and [S IV].
	
	\subsection{Results}

	\subsubsection{Cluster Evolution}
	
	\begin{figure*}
		\begin{center}
			\subfloat[][]{\includegraphics[width=0.75\textwidth]{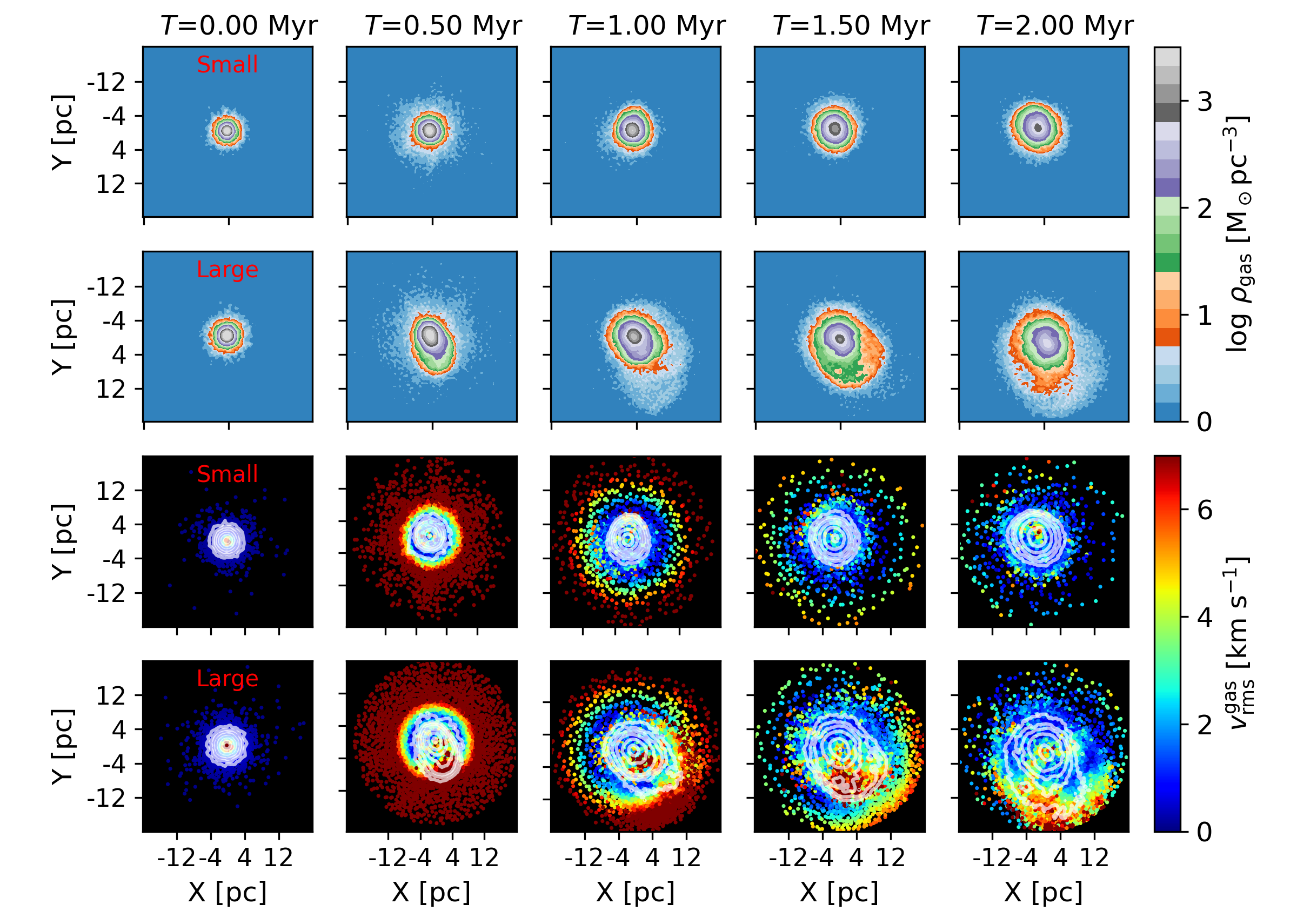}}
			\hfill
			\subfloat[][]{\includegraphics[width=0.75\textwidth]{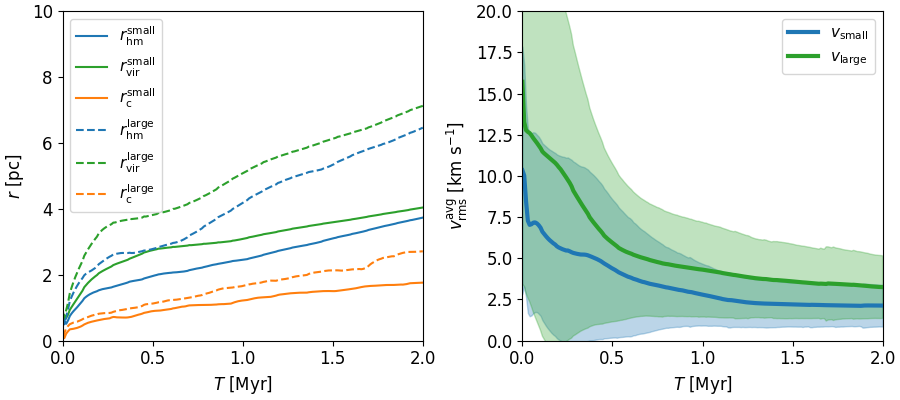}}
			\caption{Results of cluster evolution for the small and large clusters in a 10pc slice taken at the center of the Z axis at multiple timesteps (each column shows a different timestep). (a) Top two rows show the density structure (upper: small cluster, lower: large cluster). Bottom two rows show the average rms velocity structure with an overlay of the density structure contours (upper: small cluster, lower: large cluster). (b) Left: shows the half-mass, virial and core radii over time for both clusters. Right: shows the the average rms velocity with standard deviation wings over time for both clusters.  }
			\label{fig:cluster_evolution}
		\end{center}
	\end{figure*}
	
	The first step of the modelling was to follow the evolution of the clusters before the collision; the results are shown in Figure \ref{fig:cluster_evolution}.  We find that the stellar winds were the most significant factor in shaping the clusters, which is consistent with the results of simulations run by \citep{pelupessy2012evolution}; they blew away some of the gas particles, resulting in an overall increase of the clusters radii, and a decrease in the gas density, temperature and average rms velocity.  This occurred in both clusters and significantly altered the shape of the larger cluster.  The outer regions of the larger cluster are visibly distorted; the higher density small cluster remains mostly spherically symmetric. As the initial model was very general (with the one major omission of not yet including magnetic field effects) we expect these findings to be valid for typical clusters. 
	\subsubsection{Cluster Collision}
	\begin{figure*}
		\begin{center}
			\includegraphics[width=0.75\textwidth]{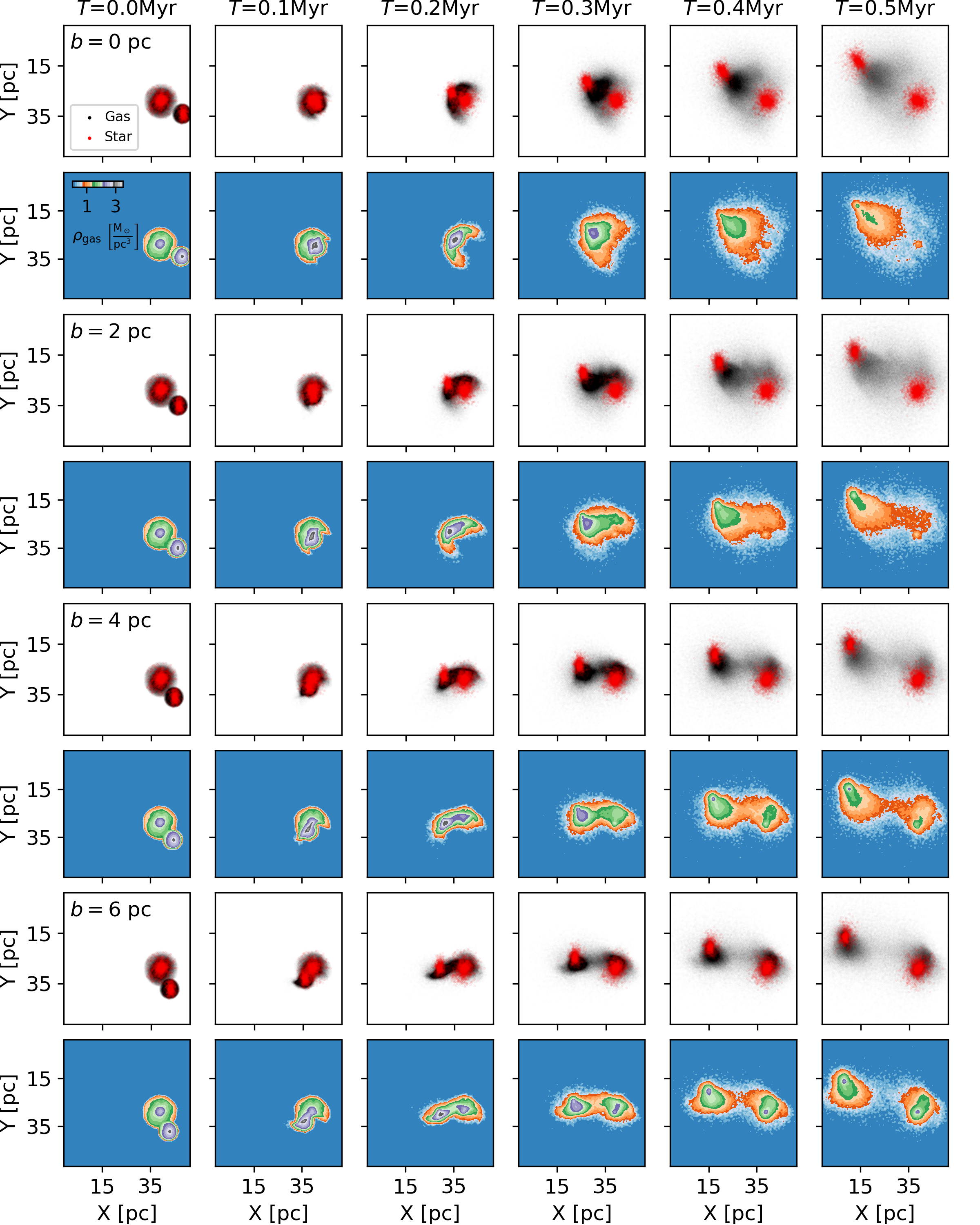}
			\caption{Cluster collision density comparison for two impact parameters at multiple timesteps; each column is a different timestep.  Every pair of rows shows the collision results for a different impact parameter ranging from $b=0$ to 6pc (where the density scale is [0, 3.5]$\msun$ pc$^{-3}$ ). The odd rows show the distribution of the gas SPH particles and the stars as black and red dots respectively. The even rows show the average gas density in a 10pc slice along the Z axis. }
			\label{fig:collision_density_comparison}
		\end{center}
	\end{figure*}
	
	\begin{figure*}
		\begin{center}
			\includegraphics[width=0.75\textwidth]{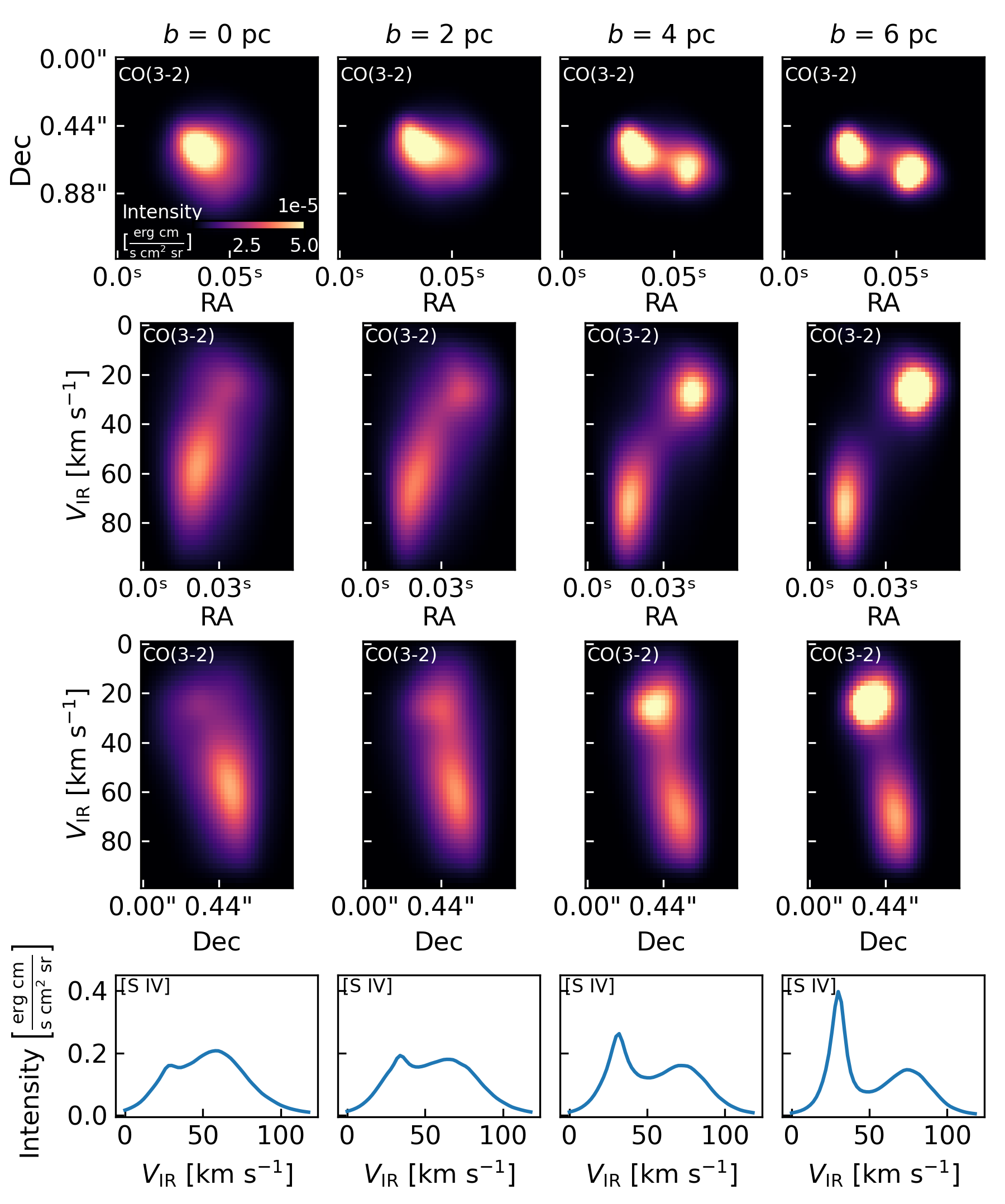}
			\caption{Results of the line profile simulations for the CO(3-2) and [S IV] lines at a fixed observation inclination angle of 45° at $T=0.5$Myr, where each column shows the results for a different impact parameter. The CO(3-2) moment map and PVDs were convolved with the beam size of the observations (the intensity scale was [1,5]$\times 10^{-5}$ \flux). Top row shows the 0\textsuperscript{th} moment maps, middle rows show the PVDs along the Dec and RA axes, and the bottom row shows the line profile for [S IV]. }
			\label{fig:rad_simulation_results}
		\end{center}
	\end{figure*}
	
	The density and distribution of gas and stars found from the simulations is shown in Figure \ref{fig:collision_density_comparison} for impact parameters of 0-6 pc in the timespan of 0 to 0.5Myr. The first point to note in this figure is how the impact parameter determines the shape of the gas distribution  In head-on collisions, the small cluster drags gas from the large cluster and spreads it out in a thick wedge-shaped tail. The stellar distribution also changes; the small cluster's distribution becomes more ellipsoidal, while the larger cluster's stays spherical and spreads to a slightly larger radius. In the grazing ($b=4$pc)  collision, both clusters maintain their overall cohesiveness but each cluster's gas spreads out across the growing divide between the two, resulting in a lower average density. This pattern is repeated at all impact parameters $>2$pc,  as can be seen in the Appendix Figure. The stellar distributions do not change drastically.  
	
\begin{figure*}
		\begin{center}
			\includegraphics[width=0.75\textwidth]{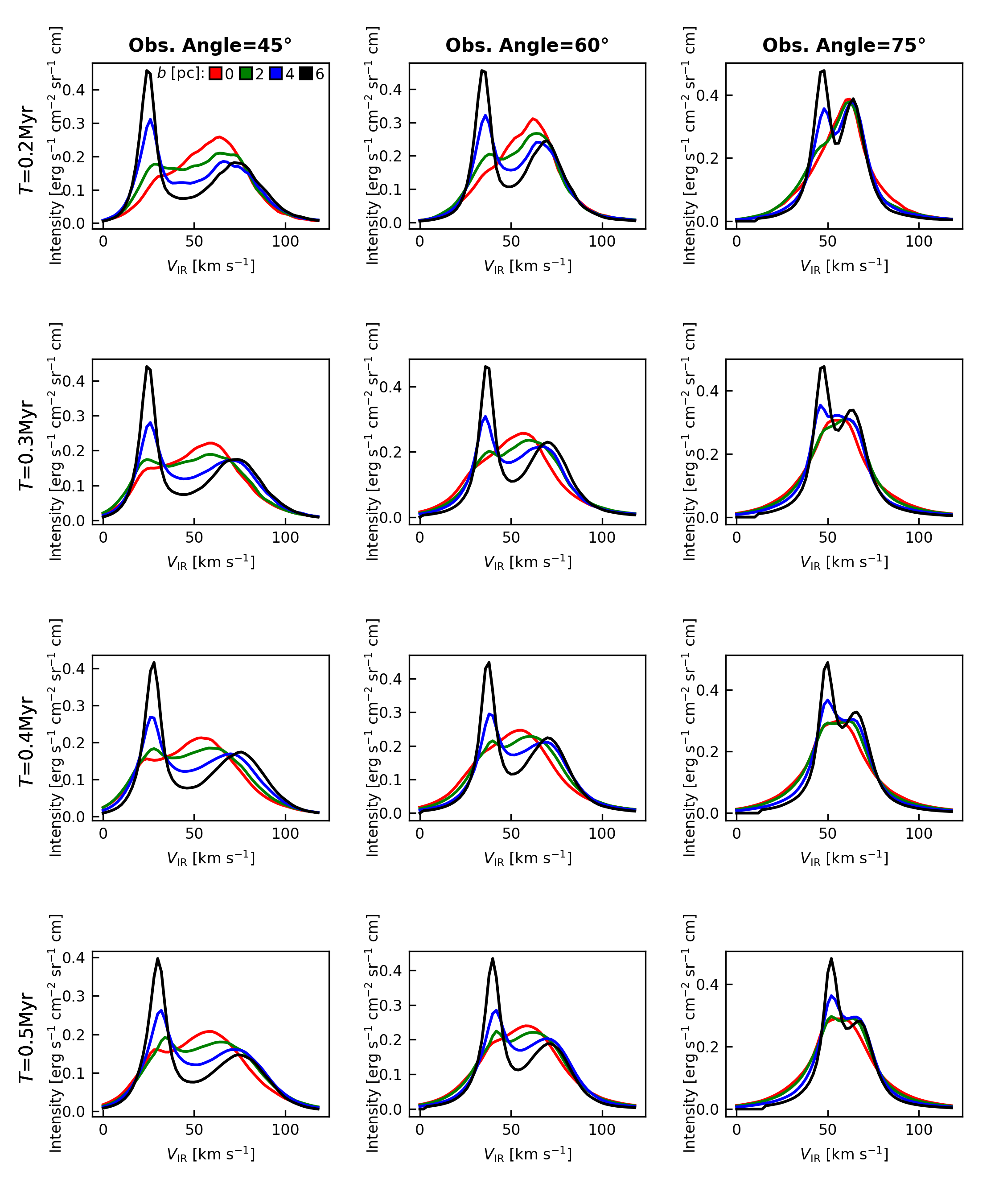}
			\caption{[S IV] line profile simulation results for multiple impact parameters, where each row corresponds to a different timestep and each column corresponds to a different observation inclination angle.  }
			\label{fig:appendix_rad_simulation_results_line_profile}
		\end{center}
	\end{figure*}

	Comparing the CO 0\textsuperscript{th} moment map and PVDs at different impact parameters displayed in Figure \ref{fig:rad_simulation_results}, shows that as the impact parameter increases the larger cluster becomes more distinct from the small one both spatially and kinematically. This is related to the gas distribution change pattern in the density distribution seen in Figure \ref{fig:collision_density_comparison}. As the impact parameter increases, less gas is stripped by the collision, and the clusters become deformed but still remain cohesive. If we consider the kinematic distribution seen in the CO PVDs and [S IV] line profile, the kinematic distance between the two clusters remains constant for increasing impact parameters and at different timesteps. The ratio between the two peaks is the major change, which increasingly favors the small cluster as the impact parameter rises. The other important factor is the inclination observation angle, which changes the spatial and kinematic distance between the two peaks as can be seen in Figure \ref{fig:appendix_rad_simulation_results_line_profile}. When comparing the CO and [S IV] observations to all the simulation results, specifically in terms of the the overall shape of the two peaks, the spatial and kinematic distance between them and the ratio between the two peaks in the line profile. We determine that the best match to the CO and [S IV] observations occurs for the $b=4$pc impact parameter, at the $T=0.5$Myr timestep, when observed at a 45° inclination angle. We note that while the simulations were made to model mainly the ionized gas distribution, the large scale behaviour of gas in these conditions, in terms of the spatial distribution and kinematics, would be very similar in the CO and [S IV] as they roughly trace the same gas mass.
	
	\section{Discussion}
	
	\begin{figure}
		\begin{center}
			\includegraphics[width=1\columnwidth]{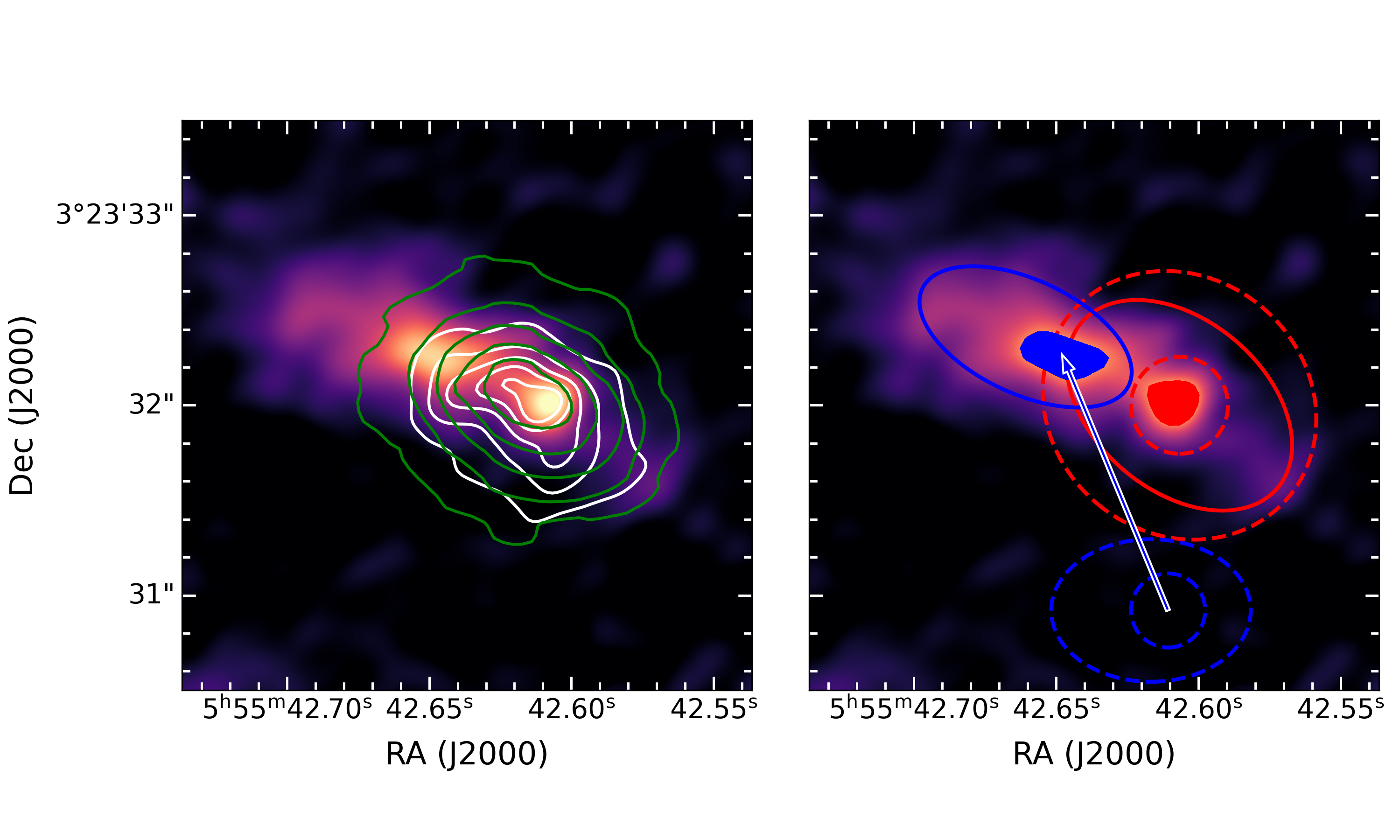}
			\caption{Left: The CO(3-2) MOM0 map overlaid with contours of the naturally weighted 15GHz map   in white and  the   [S IV] MOM0 in green. 
			  Right: The  CO(3-2) MOM0 map with rough approximations of the sizes and shapes of the small (blue) and large (red) clusters and their cores.  They are shown before and after the collision as dotted and solid lines respectively. The blue arrow shows the possible trajectory of the small cluster as it collides with the large cluster. }
			\label{fig:collision_illustration}
		\end{center}
	\end{figure}
	
	\subsection{Clusters in II Zw 40 as Modelled: Positions, Trajectories, and Possible Future Development }
		The left subfigure in Figure \ref{fig:collision_illustration} shows the CO integrated line intensity or moment 0 (MOM0) map   overlaid with the contours of the [S IV] and radio observations.    Comparing to the best matched simulation results in Figure \ref{fig:collision_density_comparison}, and allowing for the slightly ellipsoidal shape of the observed CO clouds, we identify the current positions of the clusters in the CO observations and their recent trajectory and shown this in the right subfigure of Figure \ref{fig:collision_illustration}.   The small cluster coincides roughly with the second CO peak (W2 in Figure \ref{fig:co_profiles}) and the large cluster with the main radio peak.  This reflects the cluster densities; the small cluster is denser even before the collision and the discrepancy increases in the interaction.   The ionized gas traced by radio and [S IV] emission peaks close  to the large cluster  and extends into the "bridge" area in the CO clouds (W3-W4 in  Figure \ref{fig:co_profiles} ), consistent with greater number of stars and higher ionization rate of the large cluster.  
		The  trajectory deduced for the smaller cluster, shown by the blue arrow, allows us to reconstruct the history of the collision. The smaller cluster grazes the large cluster, this collision lowering the average density of the large cluster and deforming the gas distribution. The smaller cluster, which is more compact and has a higher average density (see Figure \ref{fig:cluster_evolution}),  also deforms slightly but less than the large cluster.      The ionization observed in the "bridge"  raises the interesting possibility that a new SSC could form in the filament/band formed by the gas exchange between the clusters. This filament/band is at high pressure and density and as critical line mass depends on the width (\citet{Hacar2002}) it may fragment easily; stars can form in such conditions rapidly, matching the simulation timespan of a few hundred Kyr.   In this grazing collision, however, large shear forces are generated that may work against filament collapse and fragmentation.  The clusters in this collision are not gravitationally bound to each other;  if no new SSC forms  the bridge area between the clusters (W3)  will thin out until the two clusters separates completely. 
		
		Finally, we note that based on the [S IV] and CO distribution seen in Figure \ref{fig:collision_illustration} the small cluster appears to lie on the outer region of the [S IV] distribution, which could indicate that it contains a much smaller amount of stars that were formed prior to collision. We believe that this is due to the small cluster ionizing less of the surrounding gas due to its higher density, combined with the lower resolution of the [S IV] observations compared to the CO, resulting in the [S IV] peak being closer to the large cluster core. Alternatively, this could be due to the small cluster clump containing no stars prior to the collision, so that only the large clump contained already formed stars. In this scenario, the small clump's collision distorts the large cluster and reduces its density, so that more of the gas near the large cluster becomes ionized. A third option is the formation of a new SSC in the bridge area between the two cluster cores; for this solution to work it would have to be formed relatively closer to the large cluster's core. This discrepancy between the CO and [S IV] distributions could be solved by obtaining higher resolution observations of the ionized gas, as well as running higher resolution simulations that account for fragmentation, accretion and star formation. 
	
	\subsection{Summary and Comments}
	We have used simulations of colliding star clusters to create a model of the embedded clusters of II Zw 40 that agrees with both the spatial appearance and the kinematics of the ionized gas in the system. The most successful model is summarized in the previous section.      
	We now consider the successes and limitations of the current simulations and the questions that remain unanswered in II Zw 40. 
	
\begin{itemize}	
\item   This project concentrated on the stars and ionized gas associated directly with the clusters. The primary motivation of the models was to understand the unique spectral line profile and structure of the ionized gas on embedded source.  The models have been successful at matching the line profiles and position-velocity dependence of the [S IV] data with a simple and physically plausible scenario. 
\item  While the elongated appearance and two-component velocity structure of the molecular gas motivated the models and help establish the velocities of the components, the present simulations do not attempt to model the molecular component nor its interactions with the stars and ionized gas.   The kinematic situation of the elongated molecular clouds in II Zw 40 is intriguing. Except directly on the embedded clusters, they appear almost unaffected by the stellar winds and high gas pressures associated with the young stars.  The velocity is almost uniform across each and we do not see the shear that would show tidal effects.  What will happen as the young star clusters evolve? 
\item  The major unanswered question is whether the star clusters were formed before the collision as we posited here, or are a result of the collision of two clouds.  As the upper range of the age of SSC-N is 3Myr \citet{van2008}, both options are at present valid and the simulations cannot at present distinguish.    The ratio of molecular gas to stellar mass in this starburst is very low; the 3.5$\times10^5$\msun of molecular gas \citet{2017ApJ...850...54C} find on 'W' and 'C' is less than the stellar mass deduced for the clusters. This argues that star formation has proceeded very rapidly and efficiently to use up most of the available gas; this may have wiped out kinematic signatures that could have been seen in the molecules.
\item   To simulate the processes of cluster formation will need more models of colliding clumps that include accretion and fragmentation, as well as the interactions of molecular gas with the stellar winds of young stars.   Further questions to investigate include the role of the magnetic field in forming the clumps and controlling the interaction, and the relation of the current starburst to the interaction history of the galaxy. 

\end{itemize} 
 Theory predicts that collisions and mergers should play a major role in the evolution of star clusters and of the clouds where they form. We have the good fortune to live at exactly the right time to see this short lived interaction stage in II Zw 40, and to have the very high spectral resolution of ALMA and TEXES data to reveal the interaction process.   These results show that simulations and models of complex systems will have their greatest power when based on and guided by high resolution data.  
 \section{Acknowledgements}
 \begin{itemize} \item This paper makes use of ALMA data from program 2013.1.00122.S.  ALMA is a partnership of ESO (representing its member states), NSF (USA) and NINS (Japan), together with NRC (Canada), NSTC and ASIAA (Taiwan), and KASI (Republic of Korea), in cooperation with the Republic of Chile. The Joint ALMA Observatory is operated by ESO, AUI/NRAO and NAOJ. 
 \item The National Radio Astronomy Observatory is a facility of the National Science Foundation operated under cooperative agreement by Associated Universities, Inc.
 \item Based on observations obtained in program GN-2017A-Q-57 at the international Gemini Observatory, a program of NSF NOIRLab, which is managed by the Association of Universities for Research in Astronomy (AURA) under a cooperative agreement with the U.S. National Science Foundation on behalf of the Gemini Observatory partnership: the U.S. National Science Foundation (United States), National Research Council (Canada), Agencia Nacional de Investigaci\'{o}n y Desarrollo (Chile), Ministerio de Ciencia, Tecnolog\'{i}a e Innovaci\'{o}n (Argentina), Minist\'{e}rio da Ci\^{e}ncia, Tecnologia, Inova\c{c}\~{o}es e Comunica\c{c}\~{o}es (Brazil), and Korea Astronomy and Space Science Institute (Republic of Korea). 
 \end{itemize}
	\onecolumn
	\begin{table}
		\caption{Observational Parameters}
		\begin{tabular}{cccccc}
			\hline\hline
			Date & Telescope & Wavelength & Beam Size & Spectral Channels & noise \\
			& & & & & Jy beam$^{-1}$ \\
			\hline
			15/10/16 & JVLA & 22 GHz &$0.12\times0.07$\arcsec  (briggs wt)& NA & $1.6\times10^{-5} $ \\
			31/10/16 & JVLA & 15 GHz &  $0.2\times0.1$\arcsec (briggs wt.) & NA & $2\times10^{-5} $ \\
			13/08/14 &  ALMA & 345 GHz & $0.34\times0.25$\arcsec & 2\kms    & $3.5\times10^{-5} $ \\
			13/12/14 &   ALMA & 345 GHz & " & "  & "\\
			15/03/17 & Gemini North  & 10.5\um & $0.3\times0.3$\arcsec & 0.95 \kms &  $5\times10^{-3}$ \\
			& & & & & \flux \\
			\hline
		\end{tabular}
	\end{table}
	\section{Data Availability}
	The data used in this paper are available to the public at the archives of the observatories where they were obtained: the radio continuum  data is on the NRAO archive, the molecular data at the ALMA archive, and the [S IV] data is at Gemini North.
	
	\twocolumn 
	
	\bibliographystyle{mnras}
	\bibliography{iizw40}
	\vskip 2in
	\section{APPENDIX}
	\subsection{[S IV] Spectral Line Profiles}

	\begin{figure}
		\begin{center}
			\includegraphics*[width=1\columnwidth]{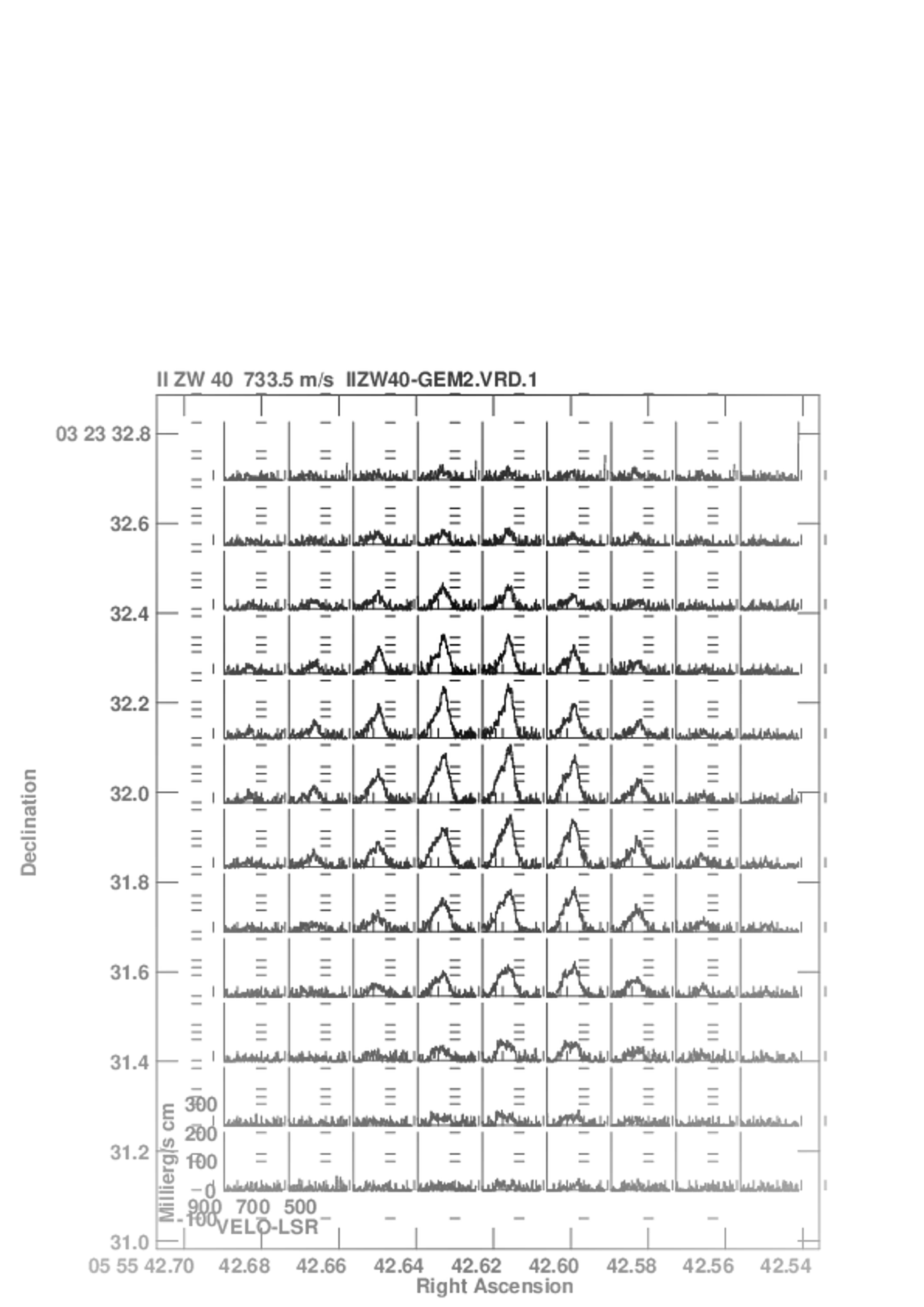}
			\caption{The [S IV] spectrum, Hanning smoothed,  in every spatial pixel in the II Zw 40 emission region.  Velocity is in \kms~.}
			\label{fig:siv_spectrum}
		\end{center}
	\end{figure}

	\begin{figure*}
		\begin{center}
		\includegraphics*[width=1\textwidth]{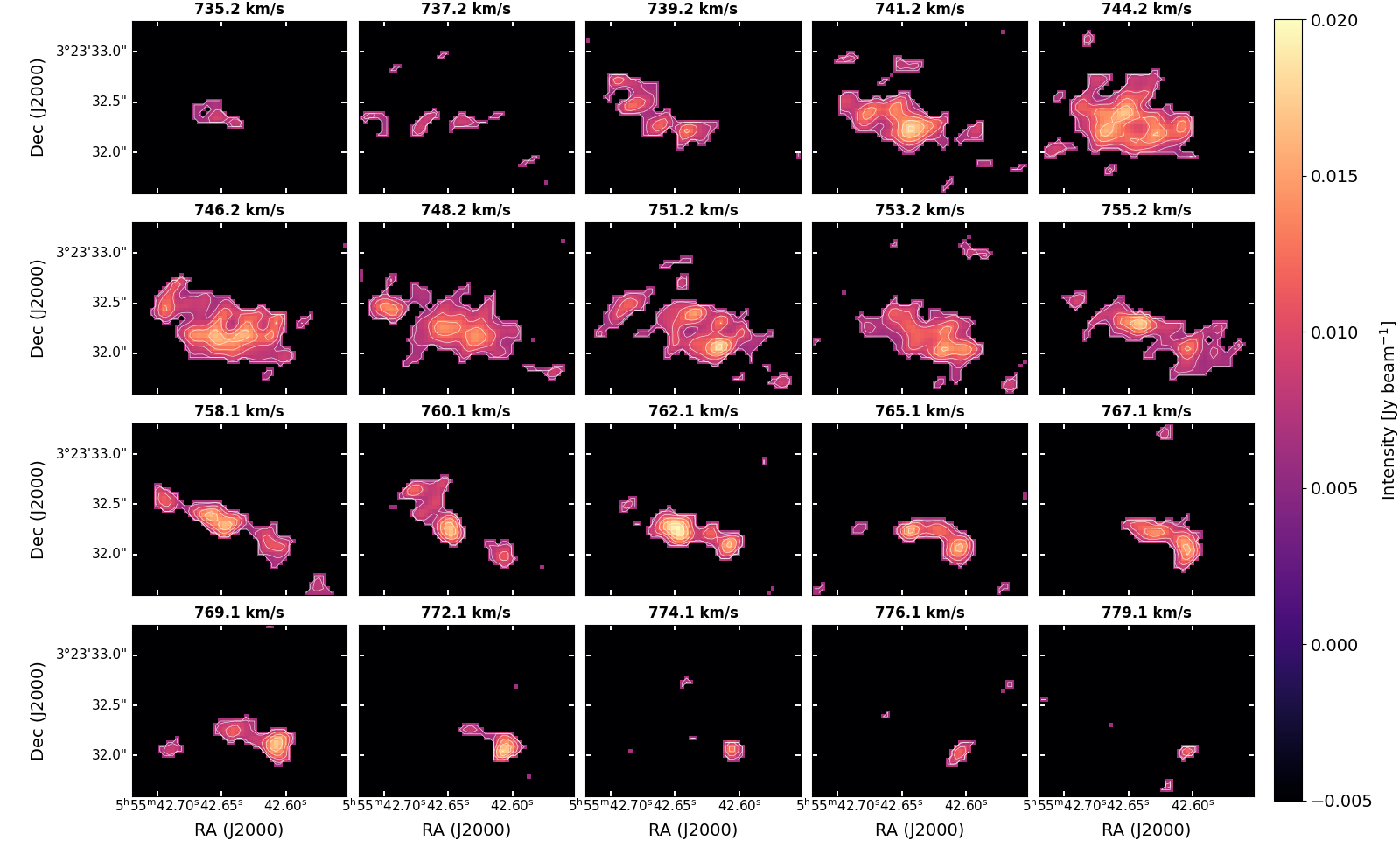}
		\caption{Channel maps of the CO(3-2) observations. The contours are at [0.005, 0.0075, 0.01, 0.0125, 0.015, 0.0175, 0.02] Jy beam$^{-1}$.}
		\label{fig:co_channel_maps}
		\end{center}
	\end{figure*}

\end{document}